\documentclass[twocolumn,showpacs,preprintnumbers,amsmath,amssymb]{revtex4}

\usepackage{graphicx}
\usepackage{amssymb}

\begin{document}
\title{Strongly Interacting Two-Dimensional Bose Gases}
\author{Li-Chung Ha,$^1$ Chen-Lung Hung,$^1$\footnotemark \footnotetext{Present address: Norman Bridge Laboratory of Physics 12-33, California Institute of Technology, Pasadena, California 91125, USA} Xibo Zhang,$^1$\footnotemark \footnotetext{Present address: JILA, National Institute of Standards and Technology and University of Colorado, Boulder, Colorado 80309-0440, USA}
 Ulrich Eismann,$^{1,2}$ Shih-Kuang Tung,$^1$ and Cheng Chin$^{1,3}$}
\address{$^1$The James Franck Institute and Department of Physics, University of Chicago, Chicago, IL 60637, USA\\
$^2$Laboratoire Kastler Brossel, ENS, UPMC, CNRS UMR 8552, 24 rue Lhomond, 75231 Paris, France\\
$^3$The Enrico Fermi Institute, University of Chicago, Chicago, IL 60637, USA}

\date{\today}

\begin{abstract}
We prepare and study strongly interacting two-dimensional Bose gases in the superfluid, the classical Berezinskii-Kosterlitz-Thouless (BKT) transition, and the vacuum-to-superfluid quantum critical regimes. A wide range of the two-body interaction strength $0.05<g<3$ is covered by tuning the scattering length and by loading the sample into an optical lattice. Based on the equations of state measurements, we extract the coupling constants as well as critical thermodynamic quantities in different regimes. In the superfluid and the BKT transition regimes, the extracted coupling constants show significant down-shifts from the mean-field and perturbation calculations when $g$ approaches or exceeds one. In the BKT and the quantum critical regimes, all measured thermodynamic quantities show logarithmic dependence on the interaction strength, a tendency confirmed by the extended classical-field and renormalization calculations.
\end{abstract}


\maketitle
Two-dimensional (2D) Bose gases are an intriguing system to study the interplay between quantum statistics, fluctuations, and interaction. For noninteracting bosons in 2D, fluctuations prevail at finite temperatures and Bose-Einstein condensation occurs only at zero temperature. The presence of interaction can drastically change the picture. With repulsive interactions, fluctuations are reduced and superfluidity emerges at finite temperature via the Berezenskii-Kosterliz-Thouless (BKT) mechanism \cite{Berezinskii72,Kosterlitz73}. Interacting Bose gases in two dimensions and BKT physics have been actively investigated in many condensed matter experiments \cite{Bishop78, Safonov98, Sebastian06, Resnick81, Coron99}. In cold atoms, the BKT transition and the suppression of fluctuations are observed based on 2D gases in the weak interaction regimes \cite{Hadzibabic06, Clade09, Tung10, Hung11}.

Extensive theoretical research on 2D Bose systems addresses the role of interactions in the superfluid phase \cite{Astra09, Popov72, Cherny01, Astra10, Mora09, Popov83} and near the BKT critical point \cite{Prokof'ev01,Prokof'ev02}. In the weak interaction regime, the classical $\phi^4$ field theory \cite{Prokof'ev01,Prokof'ev02} predicts the logarithmic corrections to the critical chemical potential $\mu_\mathrm{c}=k_\mathrm{B}T (g/\pi)\ln (13.2/g)$ and the critical density $n_\mathrm{c}=\lambda_{\mathrm{dB}}^{-2}\ln (380/g)$ for small two-body interaction strength $g<0.2$. Here $k_\mathrm{B}T$ is the thermal energy and $\lambda_{\mathrm{dB}}$ is the thermal de Broglie wavelength. The classical-field predictions are consistent with weakly interacting 2D gas experiments \cite{Kruger07,Clade09,Tung10,Hung11}.

\begin{figure}[t]
\includegraphics[width=3.4 in]{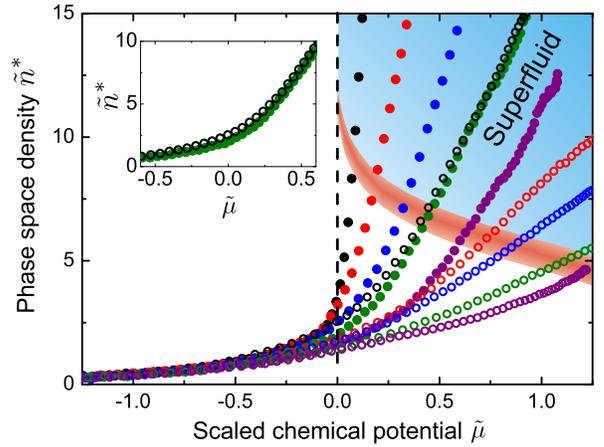}
\caption{\textbf{(color online) Equations of state for 2D Bose gases and 2D lattice gases with $\mathbf{0.05\leq g\leq 2.8}$.} The filled circles represent measurements of 2D gases with (from left to right) $g=0.05$ (black), 0.15 (red), 0.24 (blue), 0.41 (green), and 0.66 (purple).  The open circles represent measurements of 2D lattice gases with (from left to right) $g=0.45$ (black), 0.85 (red), 1.2 (blue), 1.9 (green), and 2.8 (purple). The upper blue shaded area is the superfluid regime, and the red boundary corresponds to the BKT transition regime. The black dashed line $\tilde{\mu}=0$ indicates where we evaluate the density and pressure for a vacuum-to-superfuid quantum critical gas. The inset compares the equations of state of a 2D gas and a 2D lattice gas with an almost identical $g\approx 0.4$.}\label{fig1}
\end{figure}

Intriguing dependence on the interaction strength $g$ is also predicted in the ground state properties of interacting 2D Bose gases. Popov showed that the ground state chemical potential $\mu$ deviates from the mean-field result $\mu_{\mathrm{MF}}= \hbar^2g n/m$ logarithmically \cite{Popov72}. Here, $m$ is the mass of the boson, $n$ is the density, and $2\pi\hbar$ is the Planck constant. Defining the superfluid coupling constant as $G=m/(\hbar^2\kappa)$, where $\kappa=\partial n/\partial \mu$ is the compressibility, we can summarize the perturbation expansion result of $G$ as \cite{Astra09}

\begin{equation}\label{eq:expansion}
G=\frac{g}{1+A g\ln g+Bg+Cg^2\ln g+D g^2+...},
\end{equation}

\noindent where $A=-1/4\pi$ \cite{Popov72}, $B=(\ln4-2\gamma-2)/4\pi$ \cite{Cherny01}, $C=-1/16\pi^2$\cite{Astra09}, the value of $D$ remains controversial \cite{Astra10, Mora09}, and $\gamma$ is Euler's constant. To the best of our knowledge, there is no systematic experimental study testing Eq.~(\ref{eq:expansion}).

Beyond perturbation, calculations based on the renormalized classical Ginzburg-Landau theory \cite{Subir99,Subir04} at finite temperature yield the result $G=\frac{2\pi g}{2\pi+g}$ \cite{DeriveSubir}. A recent nonperturbative renormalization-group (NPRG) calculation also provides complete thermodynamic calculations. Near the vacuum-to-superfluid quantum critical point, where the chemical potential $\mu=0$ and the temperature $T=0$, dimensionless pressure $\tilde{P}$ is approximated to be $\tilde{P}=g_2(e^{-(g/9.1) W(9.1/g)})$, where $g_2(x)=\sum_{k=1}^\infty x^k/k^2$ is the Bose function. $W(x)$ is the Lambert function satisfying $W(x)e^{W(x)} = x$, and the dimensionless density is $\tilde{n}=-\ln(1-e^{-(g/3.8) W(3.8/g)})$ \cite{PrivateRancon,Rancon12}.

In this Letter, we extend our previous work on weakly-interacting 2D Bose gases \cite{Hung11} into the strong interaction regime. We test the above theoretical predictions in different regimes (see Fig.~1) and our measurements show significant deviations from the mean-field theory as well as the logarithmic dependence on the interaction strength.

A continuous evolution of a 2D quantum gas from the weak interaction ($g\ll 1$) to the strong interaction ($g\gtrsim 1$) regime is achieved by tuning the magnetic field near a Feshbach resonance \cite{Chin10} and by combining experiments with and without an optical lattice. Optical lattices enhance the interaction strength by increasing the on site density and the effective mass $m^*$. The definition of $g$ for a 2D gas (no lattice) is given in Refs.~\cite{Petrov00,Subir06,g} and for a 2D lattice gas given in Ref.~\cite{Zhang12}. Both definitions are mutually consistent and can be connected to the 2D interaction strength $g=4\pi/|\ln na_{\mathrm{2D}}^2|$, where $a_{\mathrm{2D}}$ is the scattering length in two dimensions \cite{Schick71}.

We start our experiment by preparing a degenerate Bose gas of cesium atoms in a two-dimensional optical trap \cite{Hung11, Zhang12}. The atoms are polarized in the lowest hyperfine ground state $\left| F=3, m_F=3\right>$, where $F$ is the total angular momentum and $m_F$ is its projection. The radial and axial angular trap frequencies are $(\omega_x, \omega_y, \omega_z)=2\pi\times (8,10, 1900)$\,Hz. The sample contains $2\times 10^4$ atoms with temperature $T=13-20$\,nK, well below the excitation energy in the $z$ direction such that the sample is in the quasi-2D regime \cite{Petrov00}. We use a magnetic field to tune the atomic scattering length $a=40-580\,a_0\ll l_z$ near a low field \textit{s}-wave Feshbach resonance where scattering length crosses zero at 17 G \cite{Chin04}. Here, $a_0$ is the Bohr radius and $l_z=200$\,nm is the harmonic oscillator length in the \textit{z} direction. The corresponding interaction strength is $g=0.05-0.77$.

To further enhance the interaction, we load the 2D gas into an optical lattice. A 2D square lattice is formed with a lattice constant of $532$\,nm, and the depth is set to be $V=7.1\,E_\mathrm{R}=k_\mathrm{B}\times$ 450\,nK, where the tunneling energy is $t=k_\mathrm{B}\times$ 2.5\,nK, the effective mass is $m^*=2.9(1)\,m$, $E_\mathrm{R}$ is the recoil energy, and $k_\mathrm{B}$ is the Boltzmann constant. At this lattice depth, the system is far from the unity-filling Mott insulator phase and, for all interaction strengths we study, the ground state of the system remains in the superfluid phase. For 2D lattice gases, we can tune the interaction strength up to $g=2.8$.

To ensure thermal equilibrium, we prepare the gases at different interaction strengths by adiabatically ramping the magnetic field and the lattice potential. For all 2D gas experiments, we use a 200 ms magnetic field ramp which is slow compared to the time scale of the radial motion. For the 2D lattice experiments, we adopt an adiabatic lattice potential ramp of 400\,ms \cite{Hung10}. The magnetic field ramp is performed within the first 200\,ms of the lattice ramp. For both the 2D gas and the 2D lattice gas, we monitor the subsequent density distribution for up to 200\,ms after the ramp and observe no detectable dynamics and insignificant atom loss \cite{Loss}.

We determine the equations of state by measuring \textit{in situ} atomic density profiles based on absorption imaging with a high resolution objective (numerical aperture $= 0.5$). Imaging aberrations are carefully characterized \cite{Hung11NJP}. As a result, we achieve a spatial resolution of 1.0\,$\mu$m. The atomic density is calibrated by the number fluctuation of a normal gas \cite{Hung11}. The measured density profiles are then converted into the equation of state $n(\mu, T, g)$ based on local density approximation \cite{Ho10}, where $\mu$ and $T$ are determined by fitting the density tail \cite{Yefsah11, Hung11, Zhang12}. Note that we define the zero of the chemical potential to be the energy of the lowest available single particle state in order to compare the equations of state of both 2D gases and 2D lattice gases.

\begin{figure}[t]
\includegraphics[width=3.4 in]{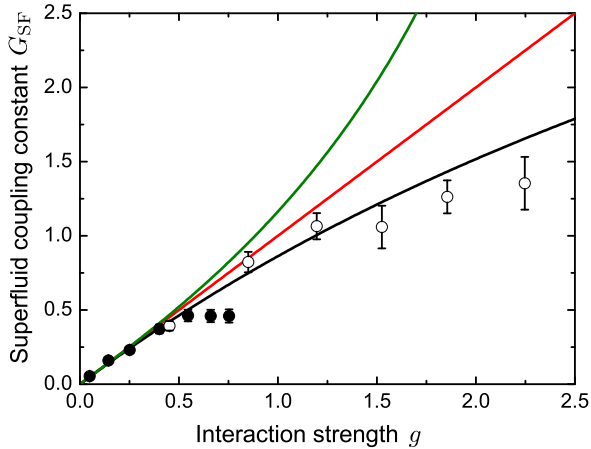}
\caption{\textbf{(color online) Coupling constant $G_{\mathrm{SF}}$ of strongly interacting 2D superfluids.} We determine $G_{\mathrm{SF}}$ by fitting the slope of the equations of state in the superfluid region for 2D gases (filled circles) and 2D lattice gases (open circles). Extensions of theoretical predictions into the strong interaction regime based on the third-order perturbation expansion \cite{Astra09} (upper green line) [see Eq.~(1)], the mean-field theory (middle red line), and the Ginzburg-Landau theory \cite{Subir04} (bottom black line) are shown for comparison. The error bars are dominated by the uncertainty of the density calibration.}\label{fig2}
\end{figure}

We plot the equations of state of 2D gases and 2D lattice gases in the dimensionless form $\tilde{n}^*(\tilde{\mu})$, where $\tilde{n}^*=n\lambda_{\mathrm{dB}}^{*2}$ is the phase space density, $\lambda_{\mathrm{dB}}^*=h/\sqrt{2\pi m^* k_\mathrm{B}T}$ is the thermal de Broglie wavelength, and $\tilde{\mu}=\mu/k_\mathrm{B}T$ is the dimensionless chemical potential \cite{Dimensionless}. For 2D gases, the effective mass is $m^*=m$. Samples of the measured equations of state are shown in Fig.~1. In particular, two equations of state with a similar $g\approx 0.4$, one from a 2D gas with $a=310\,a_0$ and one from a 2D lattice gas with $a=40\,a_0$, are compared in the inset of Fig.~1. The overall matching behavior of the two equations of state justifies our use of optical lattices to enhance the interaction. The small discrepancy near $\tilde{\mu}\approx 0$ will be discussed below.

\begin{figure}[t]
\includegraphics[width=3.4 in]{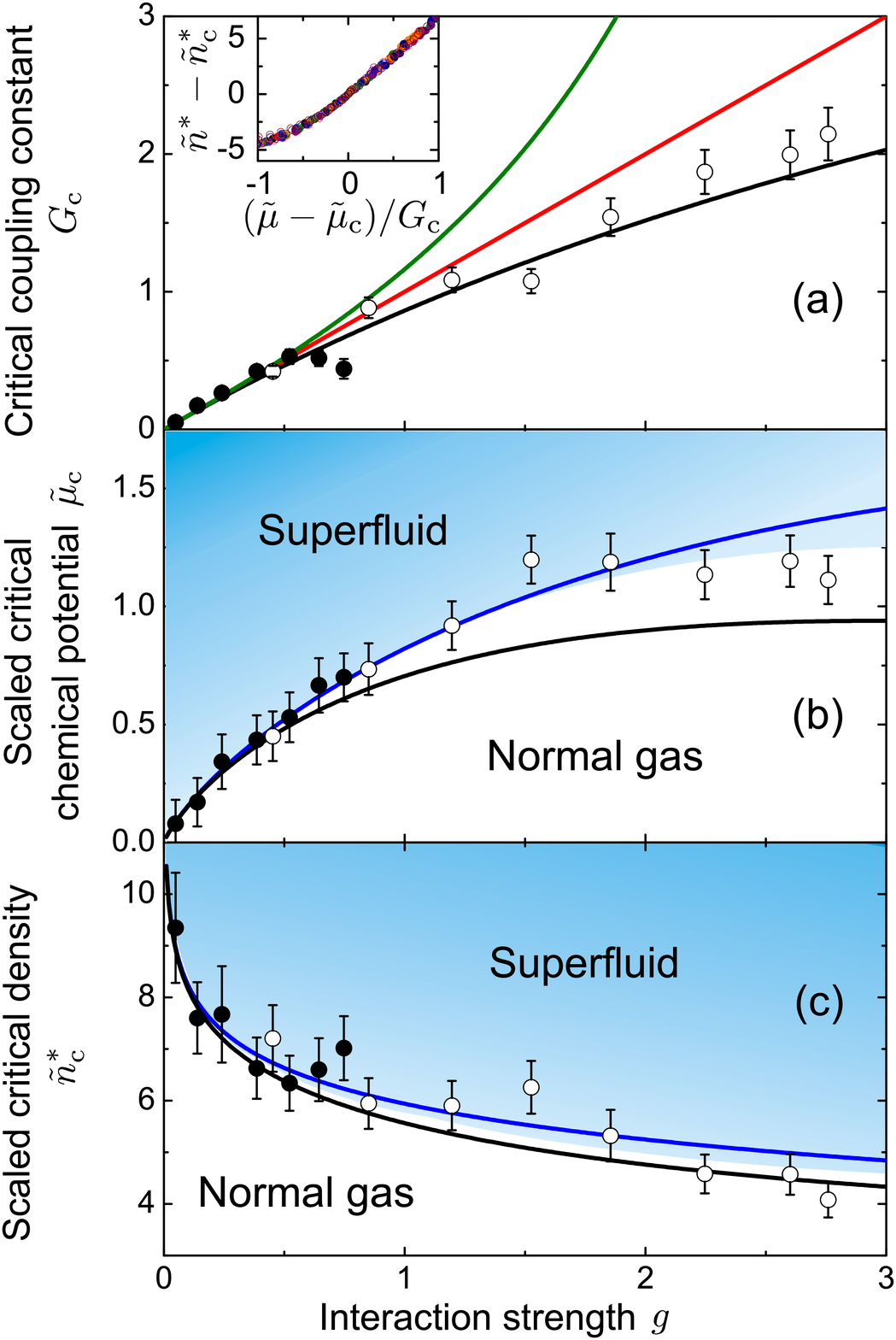}
\caption{\textbf{(color online) Critical coupling constant (a), scaled critical chemical potential (b) and scaled critical density (c) determined in the BKT transition regime.} By overlapping all scaled equations of state in the transition regime, shown in the panel (a) inset, critical parameters are determined from Eq.~(\ref{eq:uni}). The results from 2D gases (filled circles) and 2D lattice gases (open circles) are compared to the predictions from the mean-field theory (red line), the perturbation theory \cite{Astra09} (green line), the classical-field theory \cite{Prokof'ev01} (blue line), and the Ginzburg-Landau theory \cite{Subir04} (black line).}\label{fig3}
\end{figure}

In the superfluid regime, we extract the coupling constant $G_{\mathrm{SF}}=m^*/(\hbar^2\kappa)$ by evaluating the superfluid compressibility $\kappa=\partial n/\partial \mu$; see Fig.~2. The coupling constants show significant down-shifts from the mean-field prediction when the system enters the strong interaction regime. Similar tendency is also shown in the Ginzburg-Landau calculation \cite{Subir04} as well as in a recent work \cite{Mashayekhi12} which includes effective three-body interactions.

In the BKT transition regime, we use the universal critical behavior of the equations of state to determine the critical parameters \cite{Hung11, Zhang12}. By rescaling and overlapping \cite{Universality} all the equations of state in the transition regime according to

\begin{equation}\label{eq:uni}
\tilde{n}^*-\tilde{n}^*_\mathrm{c}=H(\frac{\tilde{\mu}-\tilde{\mu}_\mathrm{c}}{G_\mathrm{c}}),
\end{equation}

\noindent we obtain the critical phase space density $\tilde{n}^*_\mathrm{c}$,  the critical chemical potential $\tilde{\mu}_\mathrm{c}$, and the critical coupling constant $G_\mathrm{c}$; see Fig.~3. $H(x)$ is a generic function that describes the universal behavior near the BKT transition regime \cite{Hung11}. Remarkably, equations of state of all 2D gas and 2D lattice gas measurements overlap excellently; see Fig.~3 (a) inset.

\begin{figure}[t]
\includegraphics[width=3.4 in]{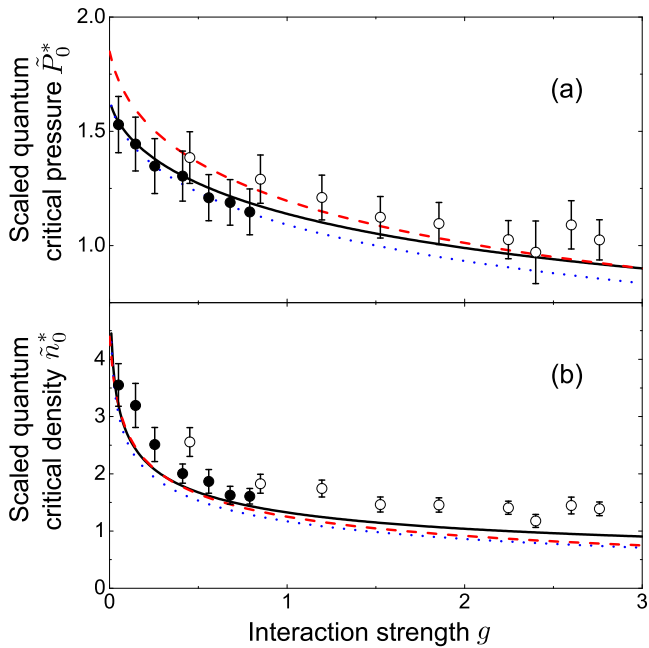}
\caption{\textbf{(color online) Pressure (a) and density (b) of a quantum critical gas at $\mathbf{\tilde{\mu}=0}$.} The measurements based on 2D gas (filled circles) and 2D lattice gas (open circles) at temperatures between $T=11-15$ nK are compared with NPRG theory \cite{Rancon12} (black solid line), mean-field theory for a 2D gas \cite{Hadzibabic08} (blue dotted line) and for a 2D lattice gas \cite{Zhang11NJP} (red dashed line) at 13\,nK.}\label{fig4}
\end{figure}

The extracted critical coupling constants $G_\mathrm{c}$ are consistently lower than the mean-field values $G=g$. On the other hand, the extracted scaled critical chemical potentials $\tilde{\mu}_\mathrm{c}$ and scaled critical densities $\tilde{n}^*_\mathrm{c}$ follow the logarithmic dependence on $g$ predicted by the classical-field calculations \cite{Prokof'ev01, Prokof'ev02, Subir04}. Our results confirm the crucial role of interactions in 2D Bose gases and suggest that the extensions of the above theories into the strong interaction regime capture the general behavior of the thermodynamic quantities.

Finally, we investigate the pressure and density in the quantum critical regime. In the lattice, atoms reach the vacuum-to-superfuid quantum critical regime when the chemical potential matches the lowest single particle state, and when the thermal energy is below the ground band bandwidth \cite{Zhang12}. We extend the definition of quantum criticality to 2D gases \cite{Subir06}. To determine the pressure, we integrate the density over the chemical potential, $\tilde{P}^*_0=\int_{-\infty}^0\tilde{n}^*\mathrm{d}\tilde{\mu}$. The extracted $\tilde{P}^*_0$ in 2D gases and 2D lattice gases are compared with the mean-field and NPRG calculations \cite{Rancon12}; see Fig.~4 (a). Here, we observe overall agreement between experiment and theories. For lattice gases, in particular, the slightly higher $\tilde{P}^*_0$ even in the weak interaction regime is discussed in Ref.~\cite{Rancon12} as the result of finite temperature effect. The densities in the quantum critical regime $\tilde{n}_0^*$ also show the expected logarithmic dependence on the interaction strength. Here, we observe small systematic deviations from the theories.

To conclude, we report the preparation and thermodynamic study of stable strongly interacting 2D gases. Dimensionless coupling constant $g$ as high as 2.8 is reached by Feshbach tuning in an optical lattice.  In the strong interaction regime, coupling constants show clear deviations from the mean-field theory. Other thermodynamic quantities in the classical and quantum critical regimes show strong dependence on $g$ and can be captured well by extensions of the classical-field theories and the NPRG calculation. Our results provide new insight into the crucial role of interactions in the thermodynamics of 2D gases as well as potential connections to other 2D condensed matter systems such as 2D Bose-Einstein condensates of spin triplets \cite{Sebastian06} and superfluid helium films \cite{Bishop78}.  Further enhancement of the interaction strength can potentially lead to crystallization of the 2D gas \cite{Xing90}. Investigation on the fluctuation and correlation of strongly interacting 2D gases will be reported elsewhere.

We thank A. Ran\c{c}on, G. E. Astrakharchik, N. Prokof'ev, E. L. Hazlett, and C. Parker for helpful discussions. We are grateful to E. L. Hazlett and C. Parker for carefully reading the manuscript. This work is supported by NSF Grant No. PHY-0747907 and under ARO Grant No. W911NF0710576 with funds from the DARPA OLE Program.


\begin{thebibliography}{99}
\bibitem{DeriveSubir} We derive $G$ starting from the scaled superfluid density $\tilde{n}=(6g/\mathcal{G}+\tilde{\mu})\pi/g+2\pi F(\mathcal{G})$; see Eq.~(A6) in Ref.~\cite{Subir04}. We then use the following approximations; $\mathcal{G}\approx 6g\exp((1+\pi/g)\tilde{\mu})/\tilde{\mu} \gg 1$ and $F(\mathcal{G})\approx (\ln\mathcal{G})/2\pi\approx (1+\frac{\pi}{g})\tilde{\mu}/2\pi$, in the superfluid limit and obtain $\tilde{n}=(1+2\pi/g)\tilde{\mu}$, or equivalently $G=2\pi g/(2\pi+g)$.
\bibitem{g} We include the logarithmic correction in the determination of $g$ according to Refs.~\cite{Petrov00,Subir06}.
\bibitem{Loss} For most experiments, atom loss is less than $5\%$. For 2D gases with $a=500$ and $580\,a_0$, $g=0.7$ and $0.8$ respectively, the loss is $10-20\%$ in the first 100\,ms.
\bibitem{Dimensionless} Throughout this Letter, we present our measurements in dimensionless form, but we do not assume scale invariance.
\bibitem{Universality} For $g$ smaller than 1.5, the range used for overlapping is $|\tilde{\mu}-\tilde{\mu}_\mathrm{c}|/G_\mathrm{c}\leq 1.5$. For $g$ larger than 1.5, it is $|\tilde{\mu}-\tilde{\mu}_\mathrm{c}|/G_\mathrm{c}\leq 1.0$.
\end{thebibliography}
\end{document}